\documentclass[preprint,amsmath,amssymb,superscriptaddress,aps, final,dvipsnames,Table]{revtex4-1}
\usepackage{rotating}
\usepackage[dvipsnames]{xcolor}
\usepackage{soul} 
\usepackage[title]{appendix}
\usepackage{ragged2e}
\usepackage{float}
\usepackage{tikz}
\usepackage{multirow}
\usepackage{chemformula}[2014/04/08]
\setchemformula{kroeger-vink=true}
\usepackage{microtype}
\usepackage{hyperref}
\hypersetup{
    colorlinks=true,
    linkcolor=blue,
    filecolor=magenta,      
    urlcolor=blue,
    pdftitle={Overleaf Example},
    pdfpagemode=FullScreen,
    }
\usepackage[labelfont=bf,justification=raggedright]{caption}

\usepackage{tabularx}
    \newcolumntype{L}{>{\raggedright\arraybackslash}X}
    \newcolumntype{C}{>{\centering\arraybackslash}X}
    \newcolumntype{R}{>{\raggedleft\arraybackslash}X}
    \usepackage{siunitx}
\usepackage{dcolumn}
\usepackage{bm}
\usepackage[]{color}
\graphicspath{ {.//} }
\definecolor{ultramarine}{RGB}{0,32,96}

\usepackage{subcaption}
\usepackage{amsmath}
\DeclareMathAlphabet{\mathcal}{OMS}{cmsy}{m}{n}
\SetMathAlphabet{\mathcal}{bold}{OMS}{cmsy}{b}{n}

\newcommand\aiidadefects{AiiDA-defects }
\newcommand\aiida{AiiDA }

\begin{document}


\title{AiiDA-defects: An automated and fully reproducible workflow for the complete characterization of defect chemistry in functional materials
}

\author{Sokseiha Muy}
\thanks{These two authors contributed equally to this work.}
\affiliation{
Theory and Simulations of Materials (THEOS) and National Centre for Computational Design and Discovery of Novel Materials (MARVEL), École Polytechnique Fédérale de Lausanne, CH-1015 Lausanne, Switzerland}%

\author{Conrad Johnston}
\thanks{These two authors contributed equally to this work.}
\affiliation{
Theory and Simulations of Materials (THEOS) and National Centre for Computational Design and Discovery of Novel Materials (MARVEL), École Polytechnique Fédérale de Lausanne, CH-1015 Lausanne, Switzerland}
\affiliation{
Catalysis Science, Pacific Northwest National Laboratory PNNL, Richland, WA 99354, United States}

\author{Nicola Marzari}
\email{nicola.marzari@epfl.ch}
\affiliation{
Theory and Simulations of Materials (THEOS) and National Centre for Computational Design and Discovery of Novel Materials (MARVEL), École Polytechnique Fédérale de Lausanne, CH-1015 Lausanne, Switzerland} 


%

\date{\today}
\vspace{10cm}
\clearpage
\newpage
\begin{abstract}

Functional materials that enable many technological applications in our everyday lives owe their unique properties to defects that are carefully engineered and incorporated into these materials during processing. However, optimizing and characterizing these defects is very challenging in practice, making computational modelling an indispensable complementary tool. We have developed an automated workflow and code to accelerate these calculations  (AiiDA-defects), which utilises the \aiida framework, a robust open-source high-throughput materials informatics infrastructure that provides workflow automation while simultaneously preserving and storing the full data provenance in a relational database that is queryable and traversable. This paper describes the design and implementation details of AiiDA-defects, the models and algorithms used, and demonstrates its use in an application to fully characterize the defect chemistry of the well known solid-state Li-ion conductors LiZnPS$_4$. We anticipate that \aiidadefects  will be useful as a tool for fully automated and reproducible defect calculations, allowing detailed defect chemistry to be obtained in a reliable and high-throughput way, and paving the way toward the generation of defects databases for accelerated materials design and discovery. 
\end{abstract}

\pacs{Valid PACS appear here}
\maketitle


\section{\label{Intro}Introduction}
Point defects in semiconductors, insulators and metals have enormous influences on the mechanical, transport, electronic, and optical properties of materials \cite{SEEBAUER200657}. 
Many materials which have found important technological applications largely owe their functionalities to the defects that were carefully selected and introduced into the pure materials during their synthesis. There is a vast literature on defect engineering to achieve specific functionalities in various functional materials, such as light emitting diodes \cite{defect_GaN_LED}, photovoltaics  \cite{perovskite_solarcell_review, photovoltaic_2019_review}, thermoelectrics \cite{thermoelectric_review}, solid-state Li-ion conductors \cite{muy_chem_review, ohno_materials_2020} and the defect centers for quantum sensing \cite{NV_center_2013, Quantum_sensing} to name just a few. 

Within the realm of computational materials science and computational chemistry, there are a growing number of packages and libraries offering some degree of workflow automation, such as ASE\cite{ase}, AFLOW\cite{aflow}, Fireworks\cite{fireworks}, QMflows\cite{qmflows} as part of domain-specific subset of the larger family of automation and parallel computing frameworks of interest to a general scientific computing audience \cite{signac,gc3pie,parsl}.
There are also a number of packages addressing different aspects of defect chemistry, especially the computation of defect formation energy corrections, and sometimes for the case of charged defects. These includes packages such as PyCDT\cite{PyCDT}, pylada-defects\cite{pylada-defects}, CoFEEE\cite{CoFFEE}, and PyDEF\cite{PyDEF}.
Typically, these employ post-, and sometimes pre-,  processing scripts to address specific aspects of defect chemistry calculations and are often designed with a particular quantum engine in mind, interacting with the chosen one either in a manual or automatic way. The number of high-quality efforts in this direction demonstrates the demand to have these complex tasks automated and abstracted away from computational scientists' workload. 

\aiida\cite{aiida, aiida-workflow-engine} is a computational infrastructure designed for  high-throughput computation and automation. Like some of the other workflow frameworks, it is an open-source project written in Python; crucially, the automation tasks are abstracted using a scalable workflow engine which runs and monitors workflow steps  concurrently, both locally and on high-performance computing resources, while simultaneously capturing all of the results and the calculation provenance in directed acyclic graphs to ensure full reproducibility of results. 
A key strength of \aiida  is its plugin system. This allows for any software or system to be interfaced with the core \aiida functionalities, with each plugin serving as an interface to the API provided by the \aiida framework.
Thanks to an enthusiastic community of plugin developers, a considerable number of scientific codes, databases and workflows are already supported including "common" workflows to calculate the same property with different quantum engines \cite{aiida-plugin-reg}. 

\aiidadefects has been developed as a plugin for \aiida with the goal of making the analysis of defect chemistry as seamless and reliable as possible, and to help accelerate the discovery and design of new engineered functional materials. 
Indeed, one of the strengths of AiiDA-defects consists in providing the automatic characterization of the complete defect chemistry of a material in a thermodynamically consistent framework, rather than just a single defect calculation, as well as providing full reproducibility of the results by leveraging the \aiida infrastructure.

To illustrate the capabilities of AiiDA-defects, we apply to study defects in the Li-ion conductor LiZnPS$_4$, which has been proposed as a solid electrolyte for the next generation of all-solid-state Li-ion batteries \cite{richards_design_2016, suzuki_synthesis_2018}.
Defects in materials used for solid-state electrolytes are particularly interesting as in their pure stoichiometric form these materials typically exhibit very low ionic conductivity. In order to reach appreciable and useful ionic conductivity, many Li-conductors require careful optimization strategies to introduce vacancies or interstitial Li into the lattice. 
This is typically achieved via aliovalent doping, where non-mobile ions are replaced with other ions of different charge, and it is assumed that charge compensation is driven by changing the concentration of Li to maintain the overall charge neutrality. However, for a given charge of the aliovalent dopant, there is no strategy to decide which ion should be chosen from the periodic table. Moreover, it is not guaranteed that changing Li concentration is always the main charge compensating mechanism, as there are potentially other defects - anti-sites, vacancies on the non-Li sites, ... - that can also lead to a charge neutral system \cite{squires_native_2020}. In order to optimize the type of defects and their concentrations, one needs to have access to the formation energies of various types of defects under given experimental conditions, such as temperature, pressure and chemical potentials of the different species. The required calculations are often resource demanding and tedious because of the large number of pre- and post-processing steps involved and the large number of calculations that are required to fully characterise the defect chemistry of the material, highlighting the need for automated workflows that can seamlessly and robustly handle these complex protocols. 

\section{\label{sec:section2} Background and methods}
A common need in the design of materials is a knowledge of the thermodynamic defects'  concentrations. In order to estimate these, one needs first to obtain the formation energy of a given defect type. The formalism for the defect formation energy (in the dilute limit) is well established \cite{zhang_chemical_1991, FreysoldtRMP}. 
Conventionally, it can be calculated using the following expression \cite{FreysoldtRMP}:

\begin{equation} \label{eq:FormationEnergy}
E^f[X^q ] = E_{tot}[X^q]-E_{tot}[bulk]-\sum_in_i\mu_i + qE_F +E_{corr}
\end{equation}
where $E^f[X^q ]$ is the formation energy of defect X in the charge state q, $E_{tot}[X^q ]$ is the energy of a supercell containing the defect (typically, the energy obtained from density functional theory (DFT) calculations), $E_{tot}[bulk]$ is the energy of the pristine supercell without the defect, $n_i$ are the number of atoms of type $i$ added ($n_i>0$) or removed ($n_i<0$) from the supercell to create the defect, $\mu_i$ is the chemical potential of atomic species $i$, $E_F$ is the Fermi level relative to the maximum of the valence band and $E_{corr}$ is the post-processing correction term added to remove the spurious long-range electrostatic interactions between the charged defect and its periodic images as well as with the uniform background charge \cite{freysoldt_PRL_2009,freysoldt_2011,Dabo2008}. 
In the \aiidadefects package, each of these terms is computed automatically in a workflow, called a ``workchain`` in AiiDA terminology, which is controlled by the \aiida workflow management engine\cite{aiida-workflow-engine, aiida}.
The standard steps of running a DFT calculation are handled seamlessly and automatically - from the preparation of input files to the retrieval of results from a remote resource - with the provenance being captured in a fully queryable database. 

In the following sections, we will describe in more detail the computational and practical considerations when computing defect formation energies, and detail the implementation of each workchain inside the package, and illustrate its use in the determination of the defect chemistry of the Li-ion conductor  LiZnPS\textsubscript{4}.

\subsection{\label{subsec:section2.1}Periodic-image correction}
In equation \ref{eq:FormationEnergy}, we stated how the defect formation energy for a given defect could be calculated. 
The most important assumption underling this equation is that the defect is in the  `infinitely dilute limit', such that the interactions between periodic images of the defect can be neglected and the defect can be considered as isolated. However, most DFT codes used periodic-boundary conditions, which leads to spurious interactions between the defect and its periodic images.
To simulate an isolated defect, one can use a supercell that is large enough such that any interaction between the defect and its periodic images has become negligible. However, a problem that remains is that generally DFT suffers from very unfavourable scaling, typically be $\mathcal{O}(n^3)$,  making this this approach impractical for realistic systems. To overcome this, it was recognised that the interaction between defects images could be thought of as arising from different contributions - that from the electronic wavefunctions overlapping, that from electrostatic interactions and that from the long-range elastic fields of the defect. Assuming these latter have become negligible in the supercell chosen, and that the electronic density of the defect state is well localised, it can be assumed that the interaction is solely from electrostatics and so model systems can be constructed with which to estimate the size of this interaction to provide an  \emph{a posteriori} correction.

The energy of such a model is given by:
\begin{equation} \label{eq:ElectrostaticEnergy}
    E = -\frac{1}{2} \int_{cell} d^3\boldsymbol{r} \rho(\boldsymbol{r}) \nabla^2\nu(\boldsymbol{r})
\end{equation}
where the electrostatic potential  $\nu(\boldsymbol{r})$ for a model system with a model charge distribution $\rho(\boldsymbol{r})$ is obtained by solving the Poisson equation \cite{Dabo2008, Dabo2011}:
\begin{equation} \label{eq:ElectrostaticPotential}
    \nabla^2\nu(\boldsymbol{r}) = - \frac{4 \pi}{\epsilon} \rho(\boldsymbol{r})
\end{equation}
where $\epsilon$ is the permittivity of the host material.

For the case of periodic boundary conditions, the potential is evaluated in reciprocal space as in:
\begin{equation} \label{eq:PoissonSolution-Reciprocal-Isotropic}
    \nu(\boldsymbol{G}) =  4 \pi \sum_{\boldsymbol{G} \neq 0} 
    \frac{\rho(\boldsymbol{G})}{\epsilon|\boldsymbol{G}^2|}
\end{equation}

To compute the electrostatic correction for a given supercell size, the difference in energy between that of model system of identical size to the DFT supercell, and an infinitely extended supercell (either extrapolated from multiple models of differing size under periodic boundary conditions, or a model with open boundary conditions) needs to be calculated. 
The remaining choice is which charge distribution to use for the model.
A number of schemes have been proposed and these can be classified into three groups \cite{Walsh2021, Walsh2020, Dabo2008}:
point countercharge corrections, such as the Makov-Payne scheme, that use a point charge in place of the charged defect \cite{Makov1995}, Gaussian countercharge corrections, that use a Gaussian type function as the model charge \cite{Dabo2008, freysoldt_PRL_2009, freysoldt_2011}, and density countercharge corrections, that use the charge density difference derived from DFT \cite{Dabo2008, Lany-Zunger}. It is also possible to directly apply these corrections in a DFT self-consistent cycle, rather than \emph{a posteriori} by adding a correction term to Kohn-Sham equation in the procedure to optimise the DFT wavefunctions \cite{Walsh2021, Suo2020, ChagasdaSilva2021}.
In AiiDA-defects we use an \emph{a posteriori} correction as this allows us to remain agnostic to the choice of any particular quantum code; in fact, recent efforts \cite{aiida-common-workflows} demonstrated how \aiida can be utilized in code-agnostic form, using the example of the computation of the equation of state to show how a common workflow can be coupled to many different codes through a set of simple  interfaces. Here we also embrace this model, and recognise that the application of a charge correction scheme \emph{a posteriori} could support any code through simple interfaces, thanks also to the fact that many solid-sate codes already have support in \aiida for the required single-point energy calculations\cite{aiida-plugin-reg}. 

 We implement and make available two options to specify the parameters of the Gaussian charge density. The default scheme is to a fit a multivariate Gaussian to the defect charge density, defined as the difference between the DFT charge density of the material with a defect in a charge state $q$ and that of the pristine bulk:
    \begin{equation}\label{eq:charge_density_difference}
        \rho_{defect}(\boldsymbol{r}) = \rho_{defect}^q(\boldsymbol{r}) -  \rho_{bulk}(\boldsymbol{r})
    \end{equation}
The probability density function for a multivariate Gaussian in 3-dimensions is given by:
\begin{equation}\label{eq:multivariate_gaussian}
    f(x) = \frac{1}{\sqrt{(2\pi)^3\det \Sigma}}\exp\left(-\frac{1}{2}(x-\mu)^T\Sigma^{-1}(x-\mu)\right)
\end{equation}
where $\Sigma$ is the symmetric positive definite covariance matrix, and $\mu$ is the mean. The fitting minimises the difference between the charge density difference, adjusting the 6 parameters of the covariance matrix and the 3-coordinate position of the mean, representing the centre of the charge density difference distribution.
The fitting automates the choice of the Gaussian width, which is a crucial consideration to enable materials and defects to be be studied in a high throughput mode.
Using a multivariate Gaussian allows for anisotropy of the model charge distribution and a better fit to the DFT charge difference. 
If a defect charge distribution is complex, e.g., displaying degenerate peaks, the charge fitting may not converge well and so the user can opt for the $2^{nd}$ option and specify the parameters of the covariance matrix directly.
This also allows for an isotropic Gaussian to be specified, similar to the Freysoldt-Neugebauer-Van de Walle type corrections \cite{freysoldt_PRL_2009, freysoldt_2011}.

In Equation \ref{eq:PoissonSolution-Reciprocal-Isotropic}, the charge density is screened by the relative permittivity $\epsilon$. 
The user may specify this value or it will be automatically calculated using denisty functional perturbation theory \cite{DFPT_Baroni}.
In practice, the full tensorial value is used and so equation \ref{eq:PoissonSolution-Reciprocal-Isotropic} is re-written as:
\begin{equation} \label{eq:PoissonSolution-Reciprocal-Anisotropic}
    \nu(\boldsymbol{G}) =  4 \pi \sum_{\boldsymbol{G} \neq 0} 
    \frac{\rho(\boldsymbol{G})}{\boldsymbol{G} \cdot \boldsymbol{\epsilon} \cdot \boldsymbol{G}}
\end{equation}
where $\boldsymbol{\epsilon}$ is a $3\times3$ matrix describing the relative permittivity in three dimensions. The user may specify a full matrix or a single value.

With these ingredients, the energy for a given size of the model system can be computed and compared to the energy extrapolated to infinity from a series of increasingly large model systems. In the case of the model charge being represented by an isotropic  Gaussian, this extrapolation can be avoided since the analytical solution of the Poisson's equation of a Gaussian charge density with open-boundary condition is known and can be used directly to obtain the isolated model energy:
\begin{equation} \label{eq:PoissonSolution-OpenBC}
    E = \frac{q^2}{2\epsilon\sigma\sqrt{\pi}}
\end{equation}
This shortcut is applied automatically if an isotropic Gaussian is supplied by the user or is found by fitting. We note in passing that these considerations apply most straightforwardly to bulk 3d materials; in the case of low-dimensional structures, extra care needs to be paid in reaching the thermodynamic limit across the poorly screened vacuum regions.

\subsection{\label{subsec:section2.2}Potential alignment}
In Section \ref{subsec:section2.1}, we showed how the interaction between periodically repeated images of the charged defect can be corrected. 
However, there is another issues arising from the implicit use of a neutralizing background when simulating a system with an overall net charge in periodic-boundary condition. This background charge compensates for the system charge to ensure overall charge neutrality, but also introduces a further spurious interaction with the defect charge \cite{freysoldt_PRL_2009, Komsa_Pasquarello}. 

\begin{figure}
\centering
\includegraphics[width=1.\linewidth]{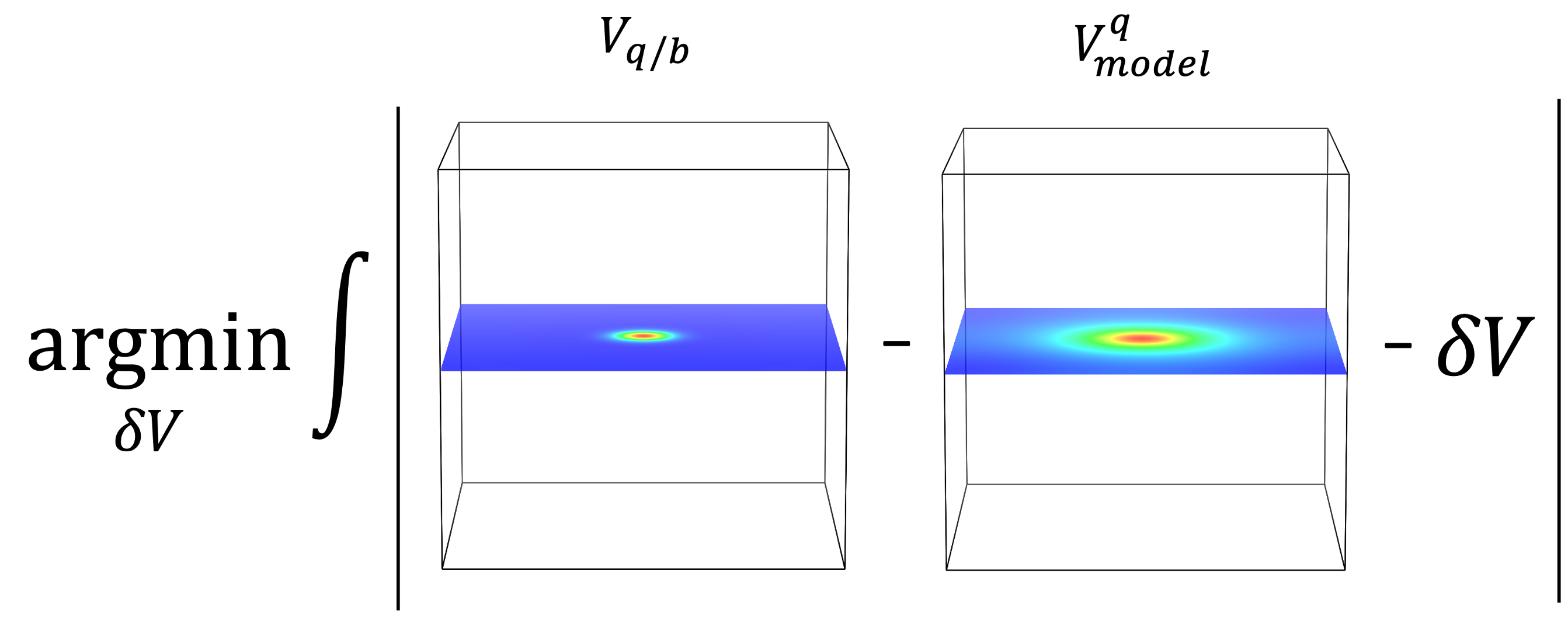}
    \caption{ In AiiDA-defects, the alignment term $\Delta V_{q/b-m}$ is obtained by computing the offset the defect short-range potential $V_{sr} \equiv V_{q/b}-V_{model}^{q}$ directly on the 3D grid and not from the planar-average. The values of the potential in the plane containing the defect site (at the center of the cell) are shown.} 
    \label{fig:potential_alignment}
\end{figure}

For an isolated defect in open-boundary condition, the electrostatic potential of defect $V_{q/b}= V_{defect}^q-V_{bulk}$ decays to zero far from the defect. However, in periodic-boundary condition, $V_{q/b}$ will plateau at a value different from zero. The offset of this plateau from zero has to be determined to reproduce the absolute position of the defect potential and constitutes the alignment term $\Delta V_{q/b-m}$ introduced by Freysoldt, Neugebauer and van de Walle\cite{freysoldt_PRL_2009, freysoldt_2011}:
\begin{equation}\label{eq:e_corr}
    E_{corr} = E_{lat} -q\Delta V_{q/b-m}
\end{equation}
where $E_{lat}$ is the correction for the spurious electrostatic interaction between the images. 
To determine the value of this plateau, one separates the defect potential $V_{q/b}$ into a long-range and a short-range contribution. The long-range part is approximated by a screened Coulomb potential created by a Gaussian charge distribution $V^q_{model}$ as described in the previous section:
\begin{equation}
    V_{q/b} \equiv V_{lr} + V_{sr} \approx V^q_{model} + V_{sr} 
\end{equation}
Since $V_{q/b}$ is known and $V^q_{model}$ can be easily computed by solving the Poisson's equation, $V_{sr}$ can be obtained and its value far from the defect is the offset $\Delta V_{q/b-m}$ that we seek to determine.


Previously, this 'alignment' was done `by eye` in one dimension, the typical procedure being that of computing a planar average over two cell vectors, so that a three dimensional potential can be reduced to a single dimension. The two potentials $V_{q/b}$ and $V^q_{model}$ are then plotted and their difference will reach a plateau at some distance sufficiently far away beyond the defect's range of interaction. The potential alignment term is then read from the value of that plateau. This procedure can be prone to error and subjectivity; to allow for automation and to avoid visual inspection, we approach this differently. Rather than computing a planar average, we retain the 3-dimensional nature of the potentials for the supercell. 
Since we want to compute the alignment at as large a distance as possible from the defect, grid points which are closer than half the distance from the defect to its closest periodic image, as per the minimum image convention, are considered to be too close to the defect center and are discard. With the remaining points, the alignment term is computed by minimizing the following objective function:
\begin{equation} \label{potential_alignment_mae}
    \Delta V_{q/b-m} = \underset{\delta V}{\arg\min} \int_\Omega \left| \left[ V_{q/b}(\boldsymbol{r})- V^q_{model}(\boldsymbol{r}) \right] - \delta V \right| d\boldsymbol{r}
\end{equation}
evaluated over the whole cell volume $\Omega$, where ${|\boldsymbol{r}-\boldsymbol{r_d}|>L/2}, \boldsymbol{r_d}$ being the position of the defect in the supercell.
This is fast and automatic, and avoids any ambiguities that may occur from averaging and fluctuations in the data (Figure  \ref{fig:potential_alignment}). If needed, one could also consider to smooth this sharp boundary, substituting for step function a smoothed one\cite{Andreussi2012}.

\subsection{\label{subsec:section2.3}Chemical potential of ions}
The chemical potential is associated with the energy cost for exchanging an atom between the materials and the surrounding thermodynamic reservoir. Its value reflects the synthesis conditions under which the materials made, and the value of the chemical potential $\mu_i$ of atom $i$ in a given compound is constrained not only by the stability of that compound but also that of all the competing phases in the phase diagram. To illustrate this idea, following Buckeridge et al. \cite{BUCKERIDGECHEMICALPOTENTIAL}, we consider a compound  $A_{m_1}B_{n_1}C_{p_1}$ in an A-B-C ternary system. We have the following conditions:
\begin{equation} \label{eq:elemental_mu}
\mu_A \leq 0; \quad \mu_B \leq 0 \quad \textrm{and} \quad \mu_C \leq 0
\end{equation}
\begin{align}
    m_1\mu_A+ n_1\mu_B+ p_1\mu_C &=\mu_{A_{m_1}B_{n_1}C_{p_1}} \nonumber \\
        &\simeq \Delta H_f(A_{m_1}B_{n_1}C_{p_1}) \label{eq:muAmBn}
\end{align}
The conditions \eqref{eq:elemental_mu} ensure that A, B and C form the compound $A_{m_1}B_{n_1}C_{p_1}$ instead of precipitating to pure element A, B and C.
Equation \eqref{eq:muAmBn} is simply the definition of the Gibbs formation energy of the compound and where we have neglected the entropic contribution and approximate $\mu_{A_{m_1}B_{n_1}C_{p_1}}$ by its formation enthalpy $\Delta H_f$. For solids, a further approximation can be made by neglecting the PV enthalpic contribution, which is typically small for condensed phases, and replacing formation enthalpies by formation energies, which are computed from DFT. 
In practice, care must be taken to ensure that these formation energies are computed with the same level of theory and numerical accuracy as those of  $E_{tot}[X^q]$ and $E_{tot}[bulk]$ in Eq. \eqref{eq:FormationEnergy}, i.e, using the same exchange-correlation functionals and plane-wave cutoffs and ensuring that k-point sampling is sufficient to converge these energy differences across different phases (e.g. metal vs insulator).

For other compounds in the A-B-C system, for example  
$A_{m_2}B_{n_2}C_{p_2}$, we will also have:
\begin{equation} 
    \label{eq:muApBq}
    m_2\mu_A+ n_2\mu_B+ p_2\mu_C  \leq \Delta H_f(A_{m_2}B_{n_2}C_{p_2})
\end{equation}
This inequality ensures that the formation of compound under consideration i.e. $A_{m_1}B_{n_1}C_{p_1}$ is favored over the formation of competing phase $A_{m_2}B_{n_2}C_{p_2}$. 
If other compounds exist in the A-B-C systems, one can generate other inequalities similar to \eqref{eq:muApBq} for each of those compounds. 
Together with Eq. \eqref{eq:elemental_mu} and \eqref{eq:muApBq}, these equations and inequalities form a system of linear constraints in $\mu_A$, $\mu_B$ and $\mu_C$ which can be solved to determine the stability region of $A_{m_1}B_{n_1}C_{p_1}$ from which the chemical potentials of A, B and C can be chosen for the calculation of the defect formation energy in Eq. \eqref{eq:FormationEnergy}. This example with ternary system can be generalized straightforwardly to any binary or multinary compound. 
As a concrete example, we will determine the stability region of Li\textsubscript{3}PO\textsubscript{4}, out of which the chemical potential of Li $\mu_{Li}$ (which is needed in eq. \eqref{eq:FormationEnergy}) can be used for the calculation of formation energy of Li-related defects, such as lithium vacancies or interstitials. Applying Eqs. \eqref{eq:elemental_mu} and \eqref{eq:muApBq} to Li\textsubscript{3}PO\textsubscript{4}, we obtain:

\begin{equation} \label{eq:mu_Li_P_O}
\mu_{Li} \leq 0; \quad \mu_P \leq 0 \quad \textrm{and} \quad \mu_O \leq 0
\end{equation}
\begin{equation} \label{eq:mu_Li3PO4}
    3\mu_{Li}+ \mu_P+ 4\mu_O = -22.164 \; eV/fu
\end{equation}
where $-22.164$ eV is the formation energy per formula unit (fu) (not to be confused with the defect formation energy) of Li$_3$PO$_4$. Note that thank to Eq. \eqref{eq:mu_Li3PO4} only two of the three chemical potentials are independent. We can therefore choose one of the elements, e.g., phosphorus as the "dependent" element whose chemical potential is fixed once the chemical potential of the "independent" elements are chosen.
In addition to Li\textsubscript{3}PO\textsubscript{4}, other stable phases in the Li-P-O system consist of {LiPO$_3$, Li$_4$P$_2$O$_7$, P$_2$O$_5$, Li$_2$O, LiO$_8$, Li$_2$O$_2$, LiP, Li$_3$P, LiP$_7$ and Li$_3$P$_7$}. For each of these competing phases, we can write down a constraint similar to Eq. \eqref{eq:muApBq} which will give respectively:
\begin{align} \label{eq:mu_other_phases}
    \mu_{Li}+ \mu_P+ 3\mu_O &\leq -13.614 \; eV/fu \\
    4\mu_{Li}+2\mu_P + 7\mu_O &\leq -35.901 \; eV/fu \\
    \dots \\
    3\mu_{Li}+ 7\mu_P \qquad  &\leq -3.605 \; eV/fu
\end{align}

\begin{figure}
\centering
\includegraphics[width=1.\linewidth]{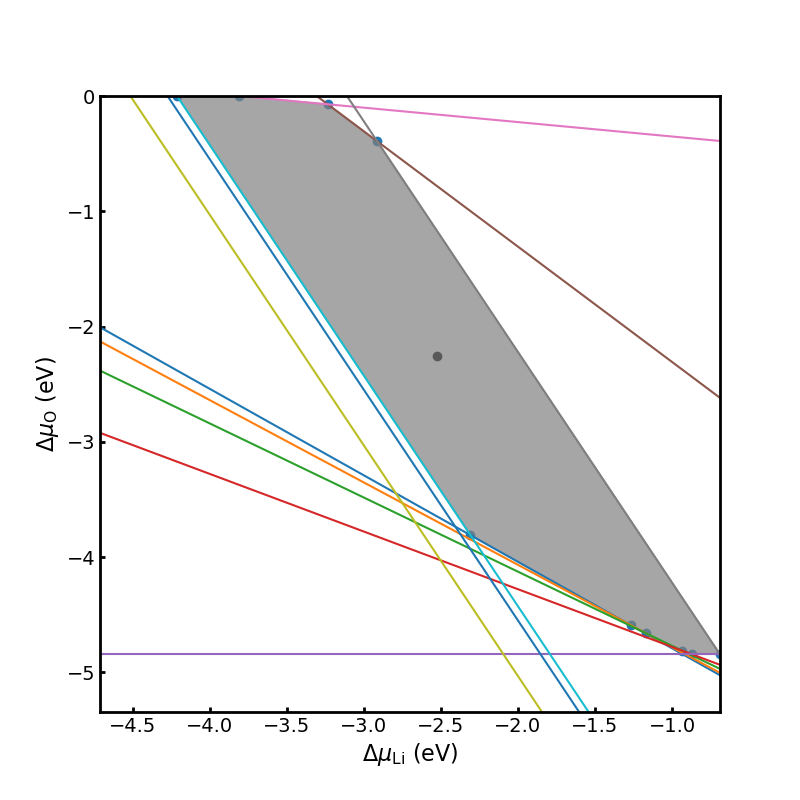}
    \caption{Computed stability region of Li$_3$PO$_4$ shown in gray as well as the centroid of the stability region (shown as the central black dot) which used by default in the calculation of the defect formation energies.} 
    \label{fig:stability_region}
\end{figure}

By solving these constraints, the stability region of Li$_3$PO$_4$ can be determined and is shown as the gray area in Figure \ref{fig:stability_region}. Any chemical potential in this region can be used to compute the defect formation energy in Eq. \eqref{eq:FormationEnergy} and by default, the chemical potential corresponding to the centroid of the stability region (the gray dot) is chosen.
Note that, so far, all the chemical potentials are \emph{relative} chemical potentials in the sense their values are referred to the reference state of each element for which the chemical potentials are set to zero. However, the chemical potentials as appear in Eq. \eqref{eq:FormationEnergy} are \emph{absolute} chemical potentials which can be obtained from the relative chemical potentials by adding the energy of the corresponding element in its reference state. 
One complication that should be mentioned is the case where the compound under consideration is predicted to be unstable; i.e., its formation energy is above the convex hull. In this instance, the formation energy of that compounds is simply shifted downward until it reaches the convex hull and a warning message is issued.

\subsection{\label{subsec:section2.4}Self-consistent Fermi level}

\begin{figure}
\includegraphics[width=1.\linewidth]{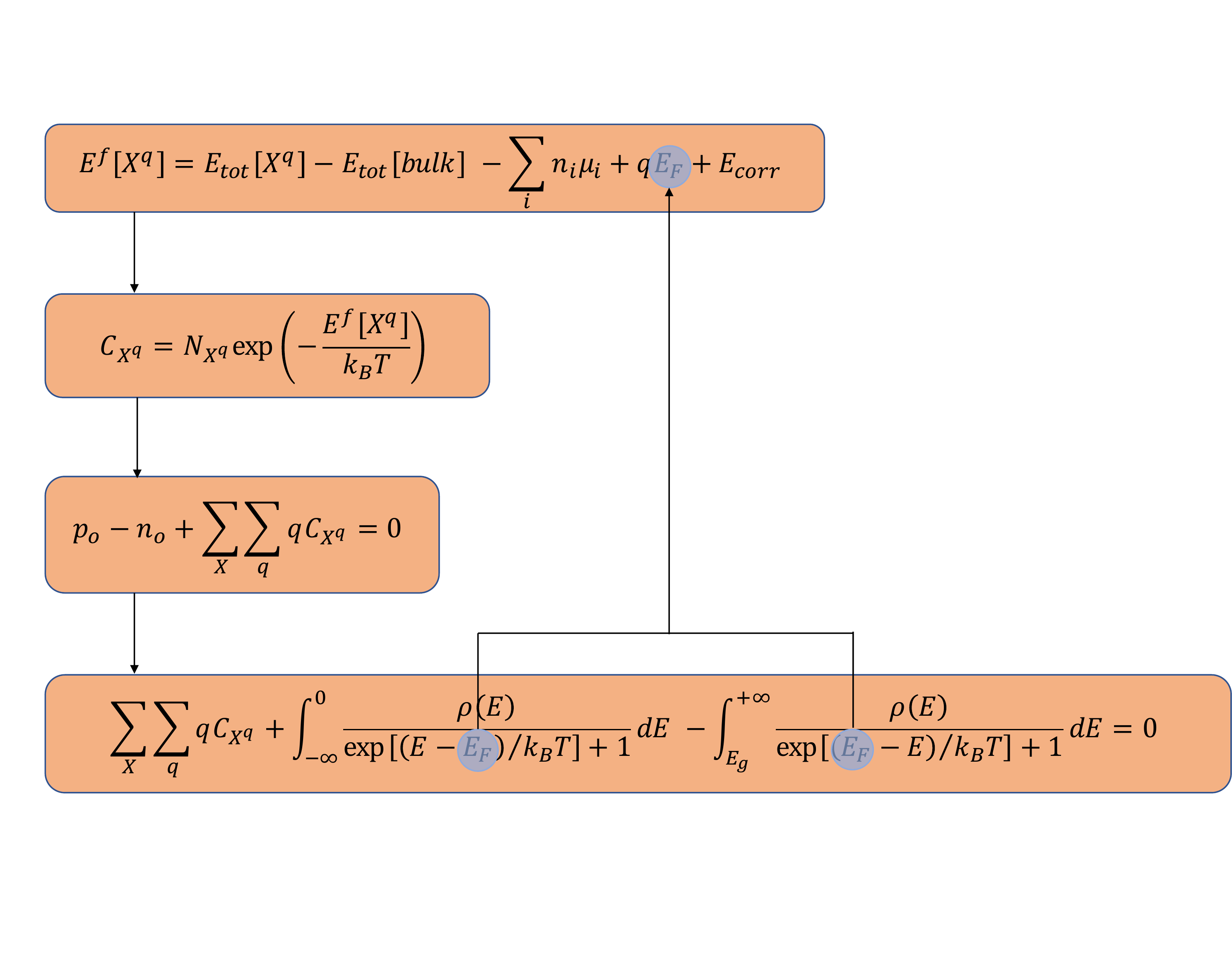}
    \caption{Algorithm used in this work to compute the Fermi level via a self-consistency cycle. Starting with an estimate of the Fermi level, the defect formation energy can be computed and, using this, a defect concentration can be estimated. By enforcing the charge neutrality condition, the concentration of electrons and holes can be computed, and from this the Fermi energy can be updated. The cycle is closed and new defect formation energies are computed. This continues until the change in $E_F$ is sufficiently small and self-consistency is achieved.} 
    \label{fig:FermiLevel}
\end{figure}

At thermodynamic equilibrium, in the presence of multiple defects with possibly multiple charge states, the Fermi level $E_F$ can be determined self-consistently by imposing a conditon of charge neutrality \cite{BUCKERIDGEFERMI}. The algorithm to compute this self-consistent Fermi level is shown schematically in Figure \ref{fig:FermiLevel}. More precisely, if we denote $n_o$ as the concentration of free electrons, $p_o$ the concentration of free holes and $C_{X^q}$ the concentration of defect $X$ in the charge state $q$, the charge neutrality condition requires that:
\begin{equation} \label{eq:ChargeNeutrality}
n_o - \sum_X\sum_q qC_{X^q} = p_o
\end{equation}
where:
\begin{equation} \label{eq:c_electron}
n_o=\int_{E_F}^{\infty}f_e(E)\rho(E)dE
\end{equation}
\begin{equation} \label{eq:c_hole}
p_o=\int_{-\infty}^{E_F}f_h(E)\rho(E)dE
\end{equation}
\begin{equation} \label{eq:c_defect}
C_{X^q}=N_{site}\exp{\left(-\frac{E^f[X^q]}{kT}\right)}
\end{equation}
and where $f_e$ is the Fermi-Dirac distribution for the electrons, $f_e(E)=[1+\exp((E-E_F)/kT)]^{-1}$, $f_h=1-f_e$ is that for the holes and $\rho(E)$ the density of states per unit volume of the pristine host. Each term $n_o$, $p_o$ and $C_{X^q}$ depends on $E_F$ and therefore the condition \eqref{eq:ChargeNeutrality} constitutes a non-linear equation in $E_F$ which can be solved numerically to determine the value of Fermi level that corresponds to a neutral system. Notice that, by using the density of states $\rho(E)$ of the perfect bulk, we implicitly used the so-called rigid band approximation which essentially states that the presence of the defects doesn't affect the electronic bandstructure of the host but simply shifts its Fermi level.

Having this algorithm implemented allows also for the possibility to fix the defect concentrations at certain values and compute the corresponding Fermi level. This feature is useful in a situation where we would like to simulate the ‘quenching’ of the system from high temperature, in which the defect concentrations are not the equilibrium concentrations at room temperature, but are rather those that are quenched from high temperature. In this scenario, one first computes the self-consistent Fermi level, and the concentration of defects and free electrons and holes self-consistently at a higher temperature and then fixes the concentration of defects but allows the Fermi level and the free electron/hole concentrations to adjust at lower temperature. This is useful for example to assess the influence of the cooling rate in high-temperature synthesis on the electronic conductivity of the materials at lower temperatures \cite{gorai_electronic_defect_2021, canepa_Mg_conductor_2017}.
Another situation in which this type of self-consistent calculation of the Fermi level can be used is to study the response of the system in terms of the change in the concentration of its native defects as a result of an \emph{effective} aliovalent doping where only the charge and the concentration of the dopants are relevant. In this approximation, the defect \ch{Li_{Mg}^x} or \ch{S_{Cl}^x} will produce exactly the same effects as a generic defect with a -1 charge at the same concentration.

\section{Package Implementation and Structure}
As previously introduced in Section \ref{Intro}, a key feature of the \aiida framework is the concept of workchains, representing a specific workflow, as a unit of computational work.
Similarly to a subroutine, a workchain accepts inputs, follows an internal logic, and returns outputs. However, unlike standard Python scripting, the inputs and outputs are captured and stored in the AiiDA database, as are the relations between these data nodes. Also, if a particular computational step in the workflow requires a calculation done by external codes, e.g., computing the DFT energy, this can also be implemented as a workchain with defined inputs and outputs via one of the numerous AiiDA plugins that interface with some of the most commonly used quantum engines \cite{aiida-plugin-reg}. 
A step in a workchain may also invoke another workchain (a child workchain) and so workchains may be specialised to a specific modular task and then nested to create complex logic. The AiiDA workflow engine is able to run all the workchains, managing each task, concurrently where appropriate, and storing intermediate results so that even if the machine running AiiDA is rebooted, the project can be resumed from the last successful step.

To best leverage this sophisticated functionality, \aiidadefects has been implemented with this modularity in mind. 
Each of the main steps in computing a defect formation energy is done  by its own workchain.
Where there is a choice of specific method for a given task, we have  implemented a parent workchain for the task which is generic, and a child workchain which implements a specific method. 
This design choice allows for any new method to be 'plugged in' by adding a workchain self-contained in a single file, and changing a minimal amount of code in the parent workchain in order to make it available as an option. 
At the heart of the package is the \emph{FormationEnergy} workchain; this workchain computes the formation energy of a given defect from the components discussed in Section \ref{Intro}.
In turn, it relies on the \emph{Correction} workchain to be able to correct the spurious electrostatic interactions. 
This workchain is in itself generic and a specific choice of correction scheme must be used. Currently, only Guassian CounterCharge method with the potential alignment functionality described in Section \ref{subsec:section2.1} and \ref{subsec:section2.2} are implemented, but thanks to the modularity of the package, other correction schemes can be easily implemented as child workchains and will be added in the future. 
The chemical potential of the defect is required to compute the formation energy, so this was implemented as a separate workchain from the \emph{FormationEnergy} workchain, and its result is passed to the \emph{FormationEnergy} workchain.
The final component is the Fermi level. However, as discussed previously, this must be self-consistent across all defects of interest to be meaningful. 
To compute a self-consistent-Fermi level, we must have knowledge of all of the defects we want to study and be able to compute their formation energies, and so it follows that the \emph{FormationEnergy} workchain must be called repeatedly by the \emph{FermiLevel} workchain. 
To manage all of these interdependent workchains, we built a master workchain called the \emph{DefectChemistry} workchain. 
While in principle a user may interact with any workchain directly, it is intended that this top-most workchain be the main interface. With this workchain, the full gamut of defect chemistry calculations can be carried out by \aiidadefects automatically.
The resulting provenance graph from an example run is shown in Figure \ref{fig:provenance} and serves to show how much automation is achieved. Each node in the provenance graph represents a data or a function performing intermediate steps in the workchain while the edges between nodes indicate whether a data is an input to a function node or is produced as an output of that function. The nodes corresponding to different child workchains are grouped and highlighted in the corresponding boxes. The provenance graph can capture and explicitly display the inter-relations between data that required or produced at any steps of the workflow.
Moreover, it should be stressed that the reproducibility is insured; without such automatic provenance capture, it would be near impossible to describe specifically what steps and calculations are behind the final data for all but the most simplistic cases. 
This shows the potential of \aiidadefects as a versatile and powerful tool for the study of defects, and how it can be used to study such defects in a high-throughput manner.
While high-throughput generally refers to high-throughput of materials to achieve some computational screening for a desired property, here we refer to it as high-throughput with respect to specific defects and charge states, allowing one to rapidly and completely characterise a material's defect chemistry. 

\section{\label{UseCases}Use cases}
\subsection{Simple defects}
\begin{figure}
\includegraphics[width=1.\linewidth]{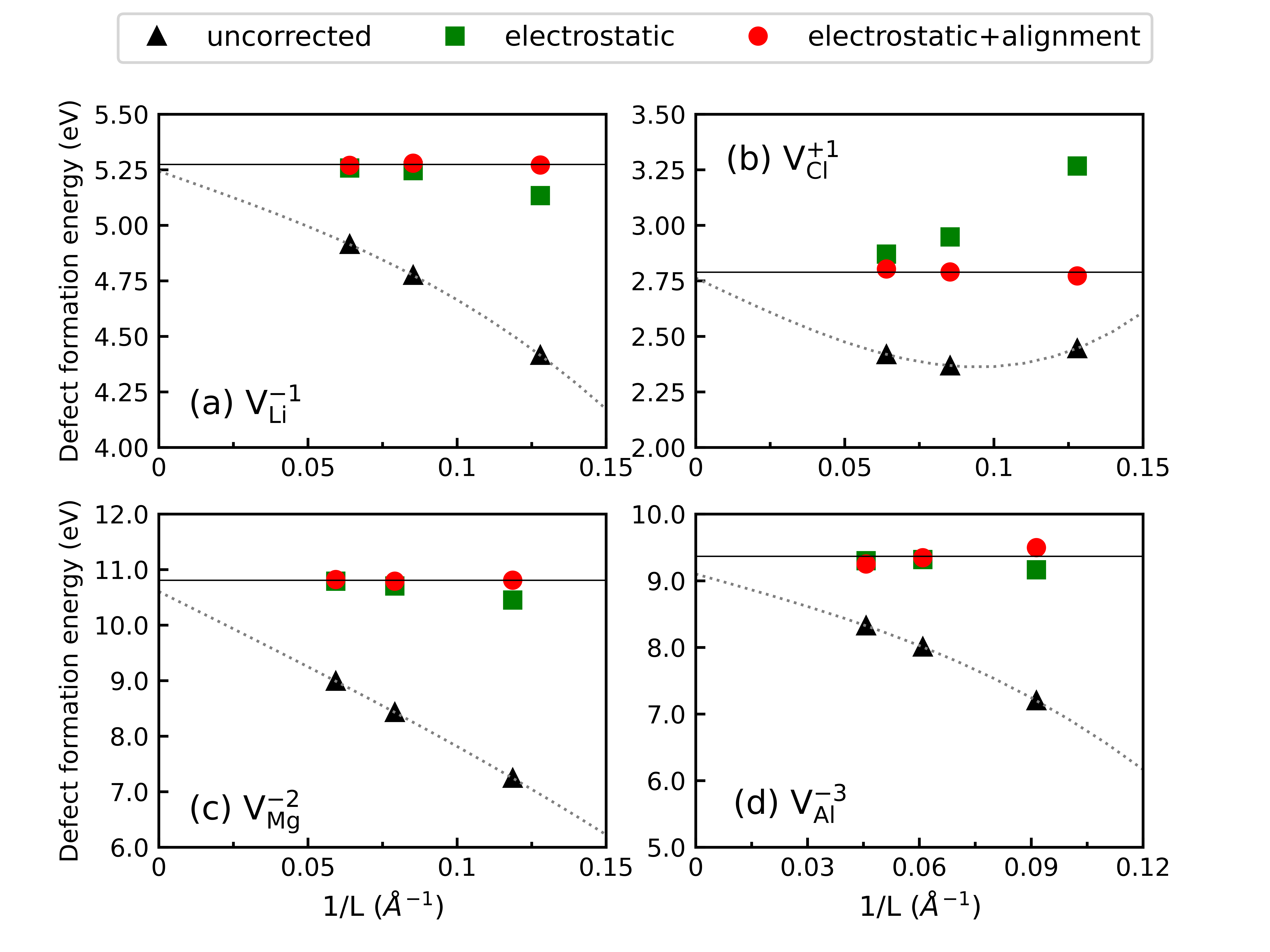}
    \caption{Formation energies of (a) $V_{Li}^{-1}$ (b) $V_{Cl}^{+1}$ in the antiperovskite Li$_3$ClO (c) $V_{Mg}^{-2}$ in MgO and (d) $V_{Al}^{-3}$ in AlP as a function of the size of the supercell.} 
    \label{fig:Ef_defect}
\end{figure}

\begin{figure*}
\centering
\includegraphics[width=1.\linewidth]{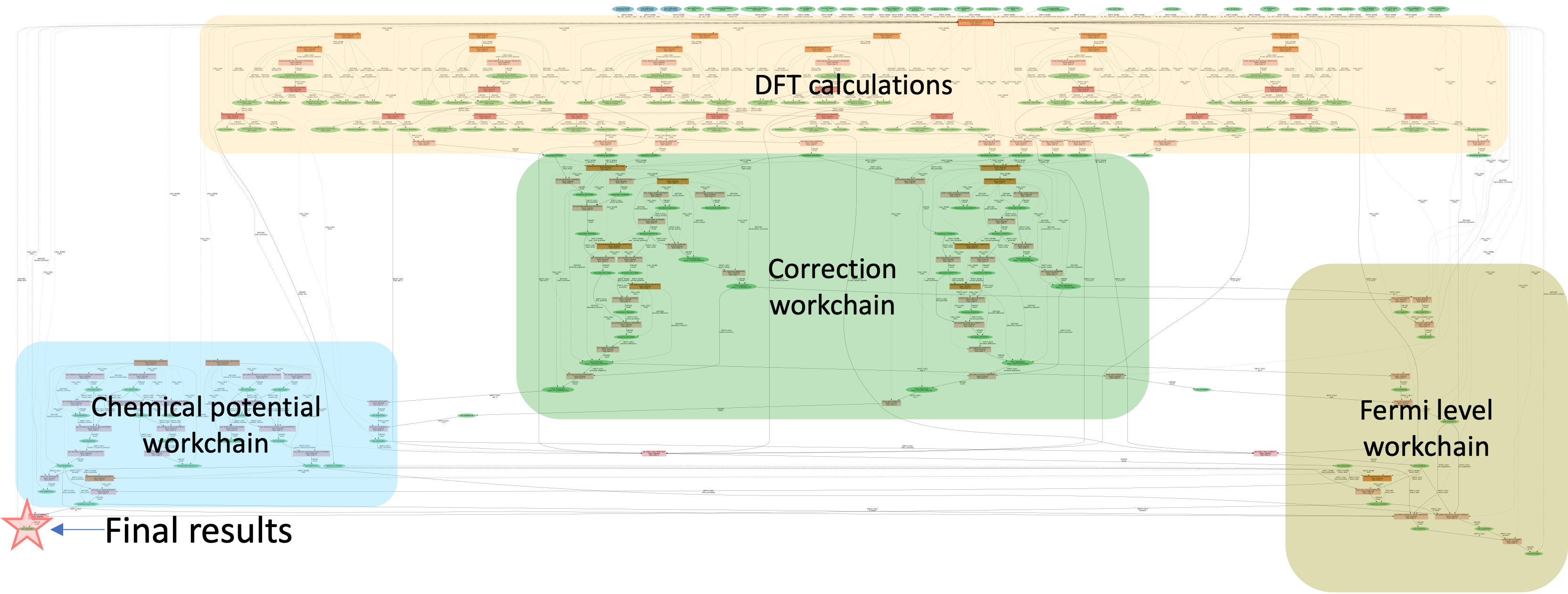}
    \caption{A provenance graph of the \emph{DefectChemistry} workchain as implemented in the \aiidadefects package. The groups of nodes belonging to different sub-workchains which compute each term in the definition of defect formation energy are shown in boxes with the corresponding name of the workchain.} 
    \label{fig:provenance}
\end{figure*}

To validate this implementation of the correction workchain, we have computed the defect formation energies of 4 different defects including $V_{Li}^{-1}$, $V_{Cl}^{+1}$ in the antiperovskite Li$_3$ClO, $V_{Mg}^{-2}$ in MgO and $V_{Al}^{-3}$ in AlP as a function of the supercell size as shown in Figure \ref{fig:Ef_defect}. As expected, the uncorrected formation energies show strong variations with the supercell size. However, upon correction, the formation energies become essentially independent of the supercell size and are all within 5\% of the their values in the dilute limit.  

\subsection{Applications}
To demonstrate the capabilities of AiiDA-Defect in a (semi) high-throughput environment, we choose to study the defect chemistry the Li-ion conductors LiZnPS\textsubscript{4}, which has been computationally predicted to exhibit extremely high Li-ion conductivity at room temperature, if the material can be synthesized with excess Li \cite{suzuki_synthesis_2018, richards_design_2016}. From the defect point of view, one way to obtain a Li-rich composition is to have partial substitution on the Zn sublattice by lithium (\ch{Li_{Zn}}) and the creation of lithium interstitial \ch{i_{Li}} to maintain charge neutrality. At thermodynamics equilibrium, whether these defects are operative or not will depend on their formation energies relative to that of other possible defects.
\aiidadefects greatly simplifies the undertaking of this kind of study.
These calculations can be conveniently carried out and the results can be easily retrieved and analyzed using the functionalities implemented in the package. 

\begin{table}[h]
\centering
\begin{tabular}{| l | c | l | c |}
\hline
\multicolumn{2}{|c|}{Native defects} & \multicolumn{2}{c|}{Extrinsic defects} \\
\hline
Defect types & Charge states & Defect types & Charge states \\
\hline
\ch{V_{Li}} & 0, -1 & \ch{Mg_{Li}} & 0, +1 \\
\ch{V_{Zn}} & 0, -2 & \ch{Mg_{Zn}} & 0 \\
\ch{V_{P}} & 0, -5 & \ch{Ca_{Li}} & 0, +1 \\
\ch{V_{S}} & 0, +2 & \ch{Ca_{Zn}} & 0 \\
\ch{Li_{i}} & 0, +1 & \ch{B_{Li}} & 0, +2\\
\ch{Zn_{i}} & 0, +2 & \ch{B_{Zn}} & 0, +1\\
\ch{Li_{Zn}} & 0, -1 & \ch{B_{P}} & 0, -2\\
\ch{Zn_{Li}} & 0, +1 & \ch{Al_{Li}} & 0, +2 \\
            &       & \ch{Al_{Zn}} & 0, +1 \\
            &       & \ch{Al_{P}} & 0, -2 \\  
            &       & \ch{Si_{P}} & 0, -1 \\ 
            &       & \ch{Sn_{P}} & 0, -1 \\  
\hline            
\end{tabular}
\caption{All native and extrinsic defects in their various charge states which were included in the calculation of defect chemistry of LiZnPS$_4$.}
\label{tab:defect_table}
\end{table}

As mentioned, in order to obtain a complete picture of defect chemistry of a material at thermodynamic equilibrium, several defects in various charge states have to be considered. For this materials, all the intrinsic (native) and extrinsic defects which are included in this study are listed in Table \ref{tab:defect_table} along with their charge states. First, we need to determine the stability region of LiZnPS$_4$ from which the chemical potential of each constitutive element can be chosen for the calculation of defect formation energy. Because LiZnPS$_4$ is a quaternary compound, its stability region is a polyhedron and a section of the polyhedron at a fixed $\Delta \mu_{Li}$ is shown in Figure. 5a. Unlike in Figure \ref{fig:stability_region}, we also directly show the concentration of the defect that we want to optimize, namely \ch{i_{Li}} in the stability region, therefore allowing a quick assessment of chemical potentials and thus, the synthesis conditions under which the concentration of the desired defects can be maximized. The concentration of various defects at room-temperature in the intrinsic regime (at chemical potentials corresponding to the centroid of the stability region, see section \ref{subsec:section2.3}) is shown in Figure 5b. One can immediately see that the dominant native defect in LiZnPS$_4$ are lithium-zinc antisite defects which are charge-compensated by the creation of interstitials lithium, leading to a lithium-excess composition as one would desire. However, this excess is very small under this condition and may not lead to any observable effects in ionic conductivity. The variation of the defect concentrations as a function of temperature can be also be obtained as shown in Figure 5c. Finally, we can also assess the dopability of this materials by examining the concentration of various aliovalent dopants as shown in Figure 5d. The most favorable extrinsic defect is \ch{Mg_{Zn}} which is an isovalent defect and will not lead to appreciable change in the concentration of lithium. \ch{Mg_{Li}} and \ch{Ca_{Li}} are thermodynamically favorable but will lead to Li antisite defects and reduce Li concentration. The most promising dopants are predicted to be \ch{Si_{P}} and \ch{Sn_{P}} which can lead to an increase of \ch{Li_{i}} via charge compensation, but their concentration at equilibrium at room temperature is predicted to be low which implies that one might need to 'quench' the defects from high temperature so that they can exist at appreciable concentrations at room temperature to produce any observable effects on the Li-ion conductivity.
This shows the robustness and user-friendliness of \aiidadefects to perform defect calculations in a high-throughput environment, and its utility to both academic and industrial users alike.

\begin{figure*}
    \centering
    \includegraphics[width=1.\linewidth]{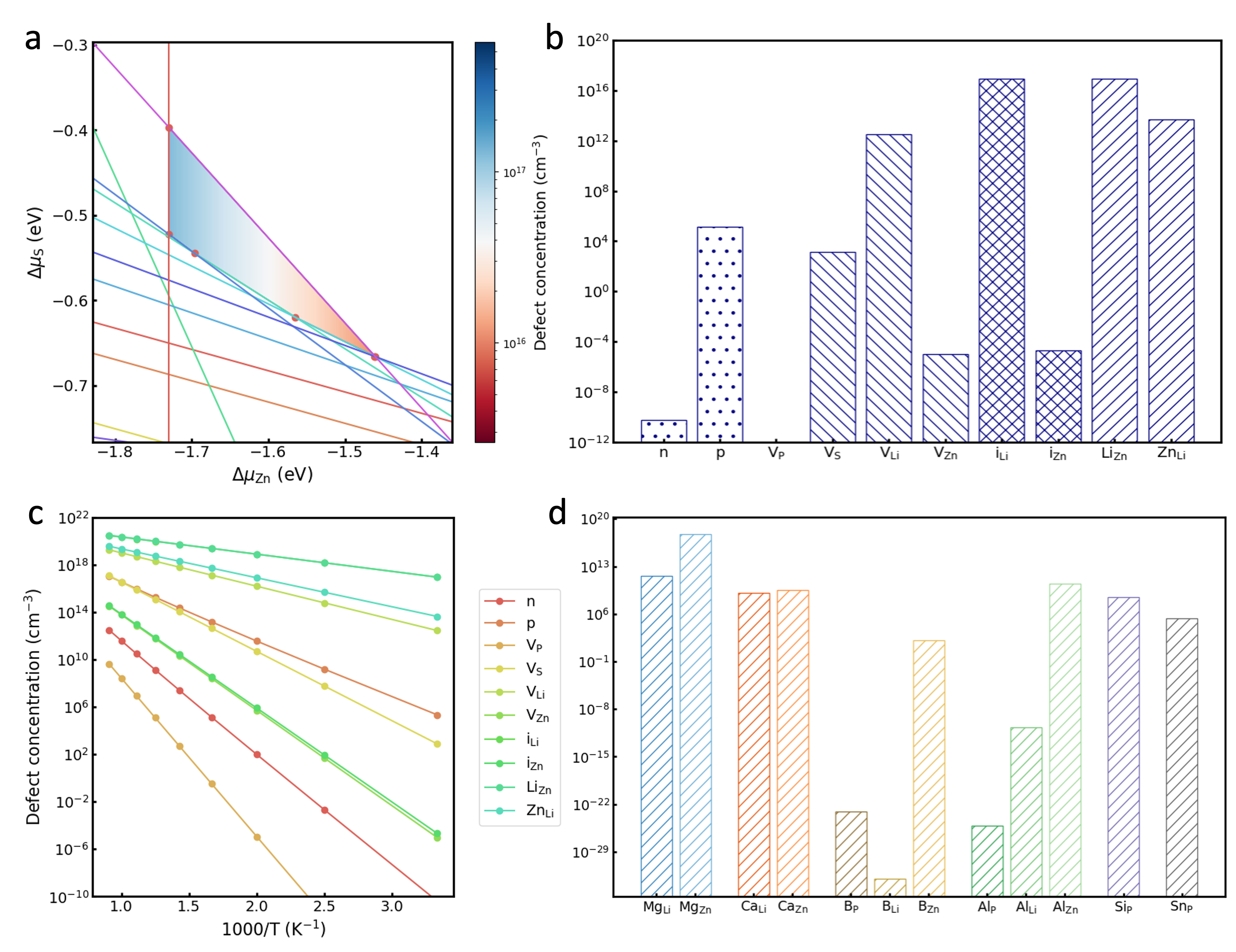}
    \caption{a) A section of stability region of LiZnPS$_4$ with the concentration of \ch{i_{Li}} directly plotted on top.
    b) Concentration of various native defects at 300K. c) Arrhenius plots of concentrations of native defects. d) Concentration of various extrinsic defects at 300K.}
    \label{fig:Examples}
\end{figure*}

\section{\label{Conclusions}Conclusions}
The accurate calculation of defect formation energies for charged systems can be challenging. We summarized the theory behind these calculations and highlighted the spurious interactions and sources of errors for these calculations when conducted within the limitations of density-functional theory calculations under periodic boundary conditions. A number of corrections and techniques have been proposed to address these technicalities, especially in the form of \emph{a posteriori} processing steps. There is an even larger menagerie of different codes and packages applying these techniques to different parts of the problem. 
These packages are very valuable, but tend to be applicable to a specific subset of use-cases or quantum codes, and may require some or even substantial manual computation and intervention.
We briefly introduced \aiida - a materials informatics platform providing a rich set of tools for automating computational workflows, distributing jobs to high performance computing resources, and for parsing and storing the results, and the associated provenance graph, in a high-performance queryable database. 
Then, we presented and described AiiDA-defects, a plugin to this framework which seeks to fully automate defect chemistry characterization and to enable this research in reliable, robust and even high-throughput modalities. In particular, our efforts have focused on identifying and solving the challenges to realizing these calculations as an automated workflow. These include the choice of charge model, the calculation of relative permittivity, the alignment of potentials, the calculation of chemical potential, the calculation and selection of an appropriate self-consistent Fermi level across a set of defects. The interaction of all of these components is taken care of automatically within a set of nested workchains. This modularity also allows for easy extension to new approximations and corrections as they are developed. 
The required DFT calculations are prepared and run automatically as needed, and like the rest of the package, the interfaces to the DFT codes are also modular, allowing for codes beyond Quantum ESPRESSO \cite{giannozzi_quantum_2009, giannozzi_advanced_2017} to be conveniently employed via existing \aiida plugins.
We showed that the correction procedures can be applied automatically to correct the formation energy of charged defects for the cases of the typical test systems. We also showed the potential of this package for a more elaborate application to the Li-ion conductors LiZnPS\textsubscript{4}, highlighting how a complete characterization of the defect chemistry can be obtained for a range of possible defects. 

\aiidadefects is powerful tool for the computational community that enables defect calculations to be performed in a robust, automated way, that is convenient and suitable for high-throughput computational studies. We hope it will be an impactful tool for new materials discovery, and a most welcome edition to the suite of community-produced packages in the \aiida ecosystem.

\begin{acknowledgments}
This work is supported by the MARVEL National Centre of Competence in Research (NCCR) funded by the Swiss National Science Foundation (grant agreement ID 51NF40-182892) and by the European Union’s Horizon 2020 research and innovation program under Grant Agreement No. 824143 (European MaX Centre of Excellence “Materials design at the Exascale”) and Grant Agreement No. 814487 (INTERSECT project). 
We thank Chiara Ricca and Ulrich Aschauer for discussions and prototype implementation ideas.
The authors also would like to thank the Swiss National Supercomputing Centre CSCS (project s1073) for providing the computational ressources and Solvay for funding this project.
We thank Arsalan Akhtar Lorenzo Bastonero, Luca Bursi, Francesco Libbi, Riccardo De Gennaro and Daniele Tomerini for useful discussions and feedback. 
\end{acknowledgments}

\textbf{Code Availability}

An open-source software implementation of AiiDA-defects is available at \url{https://github.com/epfl-theos/aiida-defects}.



\bibliography{main}

\end{document}